\documentclass[preprint,showpacs,preprintnumbers,amsmath,amssymb]{revtex4}
\usepackage{graphicx}
\usepackage{dcolumn}
\usepackage{bm}

\begin{document}
\title{End-to-end distance and contour length distribution functions of DNA helices}

\author{Marco Zoli}

\affiliation{School of Science and Technology \\  University of Camerino, I-62032 Camerino, Italy \\ marco.zoli@unicam.it}

\date{\today}

\begin{abstract}
We present a computational method to evaluate the end-to-end and the contour length distribution functions of short DNA molecules described by a mesoscopic Hamiltonian. The method generates a large statistical ensemble of possible configurations for each dimer in the sequence, selects the global equilibrium twist conformation for the molecule and determines the average base pair distances along the molecule backbone. Integrating over the base pair radial and angular fluctuations, we derive the room temperature distribution functions as a function of the sequence length. The obtained values for the most probable end-to-end distance and contour length distance, providing a measure of the global molecule size, are used to examine the DNA flexibility at short length scales. It is found that, also in molecules with less than $\sim 60$ base pairs, coiled configurations maintain a large statistical weight and, consistently, the persistence lengths may be much smaller than in kilo-base DNA.
\end{abstract}


\maketitle

\section*{I. Introduction}

Considerable research carried out over the last years has revealed that the flexibility and mechanical properties of DNA are closely related to its biological functions \cite{hart10,albu14,noy16}. 

DNA flexibility is inherent to the genome compaction in a state of negative supercoiling \cite{dekker12}, constrained within nucleosomes in eukaryotes and actively induced by gyrase enzymes in prokaryotes, which favors the initiation of replication and transcription. Bending of the molecule axis is also involved in the binding of proteins which regulate transcriptional activity and in the formation of DNA loops which bring promoter and enhancers regions in close proximity \cite{schleif92}.

Generally, DNA flexibility may not be ascribed exclusively to specific sequences. For instance, in eukaryotic chromosomes, the wrapping of about 146 base pairs around the histone octamer in a left handed superhelical path to form the nucleosome  \cite{rich97,prun98,wang12} may be considered as non-specific in terms of the sequence although heterogeneity along the path may account for the non uniform helical repeat relative to the protein surface \cite{licht14}. However, sequence specificities e.g., recurrent tracts of adenines arranged along the strand with a spacing close to the number of base pairs per helix turn (i.e., the helical repeat) of B-DNA,  may induce those curvatures of the helix axis \cite{croth90,stell13} which enhance the affinity for protein binding and ultimately activate gene transcription \cite{ohy01,wigg07,chen10}.

Analysis of looping probability induced by T4 ligase enzymes on open ends molecules showed long ago \cite{shore} that dsDNA is flexible on a scale of order of its persistence length ($l_{p}$) i.e., $\sim 150$ base pairs, although the energetic cost associated to the closure of short fragments into loops steadily increases for lengths smaller than $\sim 500$ base pairs \cite{horo}. 
These results had been explained in terms of a twisted worm-like chain (WLC) model \cite{shimada} which accounts for the observation that the ring closure probability, i.e. the $J$-factor, decreases in short molecules as a function of the length \cite{n2} and also displays a peculiar oscillating pattern. In fact, the oscillation period is a measure of the helical repeat consistently with the fact that, for DNA fragments whose number of base pairs is not a integer multiple of the helical repeat, the chain ends could be aligned only through a twist deformation all the more significant in short chains which accordingly display a decreasing $J$-factor \cite{shore1}.

However, later investigations based on ligase dependent cyclization assay \cite{widom} and fluorescence resonance energy transfer (FRET) techniques \cite{vafa,kim13}, have measured looping probabilities which, for fragments of $\sim 100$ base pairs, remain a few orders of magnitude larger than the values predicted by the WLC continuous model thus indicating that DNA molecules may maintain an intrinsic flexibility also at short length scales.  Likewise, FRET analysis of the end-to-end distance distribution combined with small-angle x-ray scattering  (SAXS) measurements  of the radius of gyration  \cite{archer08}  have suggested that the remarkable bending flexibility observed in fragments with less than $\sim 100$ base pairs may be related to a persistence length substantially smaller than the standard value of $50 nm$ for dsDNA in the kilo base pair range.  All these findings support the view that the DNA mechanical properties may depend on the fragment size \cite{gole12} and challenge the applicability of elastic rod models to those short length scales which are critical to the DNA functioning in cells \cite{maiti15}.  To address these issues  we have recently applied a mesoscopic Hamiltonian model, which treats the dsDNA at the level of the base pair \cite{io11,io14}, to the analysis of the looping probabilities in short molecules: the obtained $J$-factors fit the order of magnitude of the experimental data and reproduce the observed trend as a function of the fragment size \cite{io16b}. Importantly, radial fluctuations between the pair mates and large bending angles between adjacent base pairs have been incorporated in the computational method  thus allowing for the formation of kinks which locally unstack the helix \cite{volo08,zocchi13,kim14,ejte15}, enhance the cyclization efficiency and may reduce the persistence length  \cite{archer08,ejte12,tan15,io16a}.

Another indicator of the global flexibility of a polymer is the radial probability distribution $G(R_{e-e},\, L)$ to find the chain ends at a distance $R_{ee}$ for a given contour length $L$. Such quantity can be accessed experimentally and ( under conditions of anti-parallel alignment and twist matching of the chain ends \cite{shore1,yan15} ), in the looping limit $R_{ee} \sim \,0$, it reduces to the cyclization probability obtained e.g., for a twisted WLC model of DNA both in the case of large and  small  $L$ \cite{shimada}. 
Generally, for $G(R_{e-e},\, L)$, corrections to the Gaussian distribution of a freely jointed chain have been derived both in the case of flexible polymers \cite{dan52,stock1} with large but finite $L / l_{p}$  and for semi-flexible polymers \cite{frey96} with $L / l_{p} \sim 1$. 
While all these investigations assume the WLC model with bending elasticity as a paradigm for real polymers  \cite{wink03,ever10}, it has been recently shown that, in the case 
of short chains of nucleic acids, the WLC potential energy should incorporate also a stretch modulus term as the stretching flexibility significantly contributes to the end-to-end distribution function  \cite{tan17}.

However, in view of the fact that for chains with less than $\sim 100$ base pairs the applicability of the WLC model itself has been questioned, we feel worth to characterize the DNA equilibrium conformation and its flexibility properties following a different approach. Namely, pursuing our research line based on a mesoscopic helical model for the DNA molecule, we present in this work a calculation of the distribution functions, both for the contour length and the end-to-end distance at short length scales.  While the model can also be adapted to heterogeneous sequences by tuning the input parameters, we consider here a set of short homogeneous sequences in order to highlight the effects of the molecule size on its global flexibility. To this purpose we use a computational method which incorporates the fluctuations both in the base pair transverse displacements and in the twisting and bending variables between adjacent base pairs along the molecule stack. As radial and angular variables appear entangled in the bond lengths, the method offers a more complex and realistic description of the helical stretching flexibility than that provided by simple WLC models \cite{menon13}. Importantly, the average bond length varies in our analysis with the average twist conformation which characterizes the double helix,  the averages being carried out over the whole ensemble of base pair configurations consistent with the model potential. Thus, before proceeding to analyze the global flexibility of the molecule, one has to specify its twist conformation explicitly indicated by the average number of base pairs per helix turn. The thermodynamics provides the tools to pursue this task, selecting the equilibrium conformation and defining, with respect to the latter, both the over-twisted and the untwisted regime. In Section II, we depict our model for a helical chain and review the essential features of the Hamiltonian model and computational techniques developed over the last years.  In Section III, we determine the free energy and equilibrium twist conformations for a set of short helices. Section IV contains the analysis of the contour length and radial distribution functions. The latter results are used to extract information regarding the persistence length at short length scale in conjunction with other investigations and some available experiments. Some conclusions are drawn in Section V.

\section*{II. Model and method}

In this Section, first we depict the geometrical model for a linear helical molecule made of $N$ point-like base pairs.
Next, we summarize the key features of the coarse grained Hamiltonian for the dsDNA molecule which incorporates bending and twisting degrees of freedom. Finally, we outline the method to obtain the partition function and the statistical ensemble which defines the average equilibrium properties for the molecule conformations. For further details on the helix model, Hamiltonian and computational technique we refer to \cite{io16b}.

\subsection*{A. \, Helix representation}

\begin{figure}
\includegraphics[height=8.0cm,width=8.0cm,angle=-90]{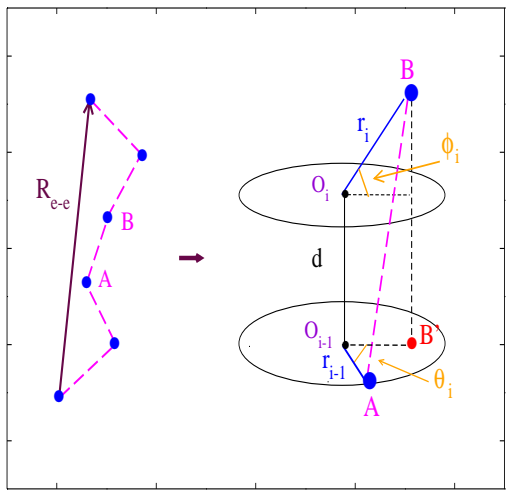}
\caption{\label{fig:1}(Color online)  
Model for an open end chain with $N$ point-like base pairs. $R_{e-e}$ is the chain end-to-end distance. The segment $\overline{AB}$, i.e. the separation between two adjacent base pairs along the molecule backbone, is given by the distance between the tips of the radial displacements $r_{i}$, $r_{i-1}$. The $r_{i}$'s represent the inter-strand fluctuational distance between the two mates of the $i-th$ base pair. They are measured with respect to the $O_i$'s which lye along the central axis of the helix. $\theta_i$ and $\phi_{i}$ are the twisting and bending angles respectively formed by adjacent $r_{i}$ and $r_{i-1}$. 
In the absence of radial fluctuations, all $r_{i}$'s would be equal to the bare helix diameter and the model would reduce to a freely jointed chain model made of $N-1$ bonds, all having length $d$.    In the absence of bending fluctuations, the model would reduce to a fixed-plane representation as depicted by the ovals in the r.h.s. drawing.
The latter also convey the idea that the $r_{i}$'s represent in-plane fluctuations.
}
\end{figure}

For the $i-th$ base pair, $(i=\,1, \,N)$, the positions of the two pair mates on complementary strands are denoted by, $r_{i}^{(1)}=\, -R_0/2 + x_{i}^{(1)}$ and $r_{i}^{(2)}=\, R_0/2 + x_{i}^{(2)}$, with $R_0$ being the inter-strands separation i.e., the bare helix diameter. $x_{i}^{(1,2)}$ represent the fluctuations of the two bases in the pair.
Accordingly, the fluctuating relative distance between the pair mates (measured with respect to the helix mid-axis) is defined by, $\, r_{i}=\,r_{i}^{(2)} - r_{i}^{(1)}$. This is the variable radial displacement visualized as in Fig.~\ref{fig:1}. 
In the absence of radial fluctuations, all $r_{i}=\,R_{0}$ and the blue dots in Fig.~\ref{fig:1} would overlap with the $O_i$'s which are arranged along the molecule central axis at a fixed rise distance $d$. This parameter corresponds to the bond length in the freely jointed chain model. 

While the $r_i$'s may also be smaller than $R_0$ due to local contractions of the helix, too large contractions are prevented by the phosphate-phosphate electrostatic repulsion between complementary strands \cite{york05} which in turn is stabilized by the counter-ion concentration in the solvent \cite{onuf16}. These effects are included in our computation by choosing suitable cutoffs on the radial fluctuations as explained below.

Successive base pair displacements, $r_{i}$ and $r_{i-1}$, along the molecule stack are twisted by the angle
$\theta_i$ and bent by the angle $\phi_{i}$.  Hence the general rise distance $\overline{d_{i,i-1}}$, that is the segment $\overline{AB}$ in Fig.~\ref{fig:1},  can be written as a function of the radial and angular variables as:

\begin{eqnarray}
& & \overline{d_{i,i-1}}^2=\, (d + r_i \sin \phi_i)^{2} + r_{i-1}^2 + (r_i \cos \phi_i)^2 -2 r_{i-1} \cdot r_i \cos \phi_i \cos \theta_i \, , 
\label{eq:0}
\end{eqnarray}

In terms of the $\overline{d_{i,i-1}}$'s, one defines the contour length $L$ and the end-to-end distance $R_{e-e}$ as:

\begin{eqnarray} 
& &L=\, \sum_{i=2}^{N} \biggl| \overline{d_{i,i-1}} \biggr| \, , \nonumber
\\ 
& &R_{e-e} =\, \biggl| \sum_{i=2}^{N}  \overline{d_{i,i-1}} \biggr|    \,.
\label{eq:0a}
\end{eqnarray}

In the following, the bare helix parameters are set to the values $R_0 = \,20 $\AA {} and $d = \, 3.4$\AA {}.

\subsection*{B. \, Hamiltonian}

We model the dsDNA molecule, made of $N$ base pairs of reduced mass $\mu$,  by a Hamiltonian that includes: i) a one particle potential $V_{1}[r_i]$ which accounts for the hydrogen bonds between complementary strands and ii) a two particles potential $V_{2}[ r_i, r_{i-1}, \phi_i, \theta_i]$ which describes the stacking forces between adjacent base pairs along the molecule backbone and carries dependence on the angular variables. Explicitly the potential terms read:

\begin{eqnarray}
& &V_{1}[r_i]=\, V_{M}[r_i] + V_{Sol}[r_i] \, , \nonumber
\\
& &V_{M}[r_i]=\, D_i \bigl[\exp(-b_i (|r_i| - R_0)) - 1 \bigr]^2  \, , \nonumber
\\
& &V_{Sol}[r_i]=\, - D_i f_s \bigl(\tanh((|r_i| - R_0)/ l_s) - 1 \bigr) \, , \nonumber
\\
& &V_{2}[ r_i, r_{i-1}, \phi_i, \theta_i]=\, K_S \cdot \bigl(1 + G_{i, i-1}\bigr) \cdot \overline{d_{i,i-1}}^2  \, , \nonumber
\\
& &G_{i, i-1}= \, \rho_{i, i-1}\exp\bigl[-\alpha_{i, i-1}(|r_i| + |r_{i-1}| - 2R_0)\bigr]  \, . \nonumber
\\ 
\label{eq:01}
\end{eqnarray}

The one particle potential contains a Morse potential $V_{M}[r_i]$, with depth $D_i$ and width $b_i$, which measures the energy associated to the transverse stretching vibrations between the pair mates from the zero level corresponding to the absence of radial fluctuations, i.e., $|r_i| = R_0$.  
$V_{M}[r_i]$ also mimics the repulsion between negative phosphate groups on complementary strands through a hard core which provides a criterion to select the range of fluctuations contributing to the partition function. Precisely, our computational method retains all base pair displacements such that $V_{M}[r_i] \leq D_i $ and consistently rejects those short displacements such that $|r_i| - R_0 < - \ln 2 / b_i$ that would bring the pair mates too close to each other. Conversely, if all displacements had to become large enough to sample the Morse plateau, i.e. $|r_i| - R_0 \gg b_i^{-1}$, then the complementary strands would go infinitely apart without any further energy cost. This situation however has no experimental counterpart as DNA in solution is not infinitely diluted. This amounts to say that the Morse potential alone cannot describe the physical case and calls for the introduction of a solvent potential, $V_{Sol}[r_i]$, to be added to $V_{M}[r_i]$. 

As discussed in refs.\cite{io11,io12}, the solvent term defined in Eq.~(\ref{eq:01}) enhances the energy threshold for base pair breaking above the Morse dissociation energy and introduces a hump which modifies the Morse plateau thus providing a barrier which accounts for those effects of strand recombination occurring in solution \cite{druk01}. Other analytical forms for $V_{Sol}[r_i]$ having the same physical properties may be used without changing quantitatively our results.

Furthermore, it is also noticed that the boundedness of the one particle potential as a whole, for large $r_{i}$, is at the origin of the well-known divergence of the partition function encountered in Hamiltonian studies of DNA denaturation \cite{zhang97}. As such divergence cannot be removed by standard methods \cite{n1}, one has to confine the phase space which the $r_{i}$'s can sample \cite{ares05}. This is consistently done by our computational method based on the path integral formalism.

The two particles potential $V_{2}[ r_i, r_{i-1}, \phi_i, \theta_i]$ depends on the elastic force constant $K_S$ and on the stacking parameters $\rho_{i, i-1} \,, \alpha_{i, i-1}$ which weigh the non-linear contribution to the intra-strand interactions.

The non-linear stacking term had been originally introduced, in a simple ladder Hamiltonian model for DNA without angular variables,  to account for the cooperative character of the base pair opening in the melting process and its observed sharpness \cite{pey93b}. 
In that model however,  whenever two adjacent bases slide far away from each other, the stacking energy becomes very large. To overcome this drawback, other models \cite{joy08} have been proposed to account for the finiteness of the stacking interaction which essentially originates from the rather rigid sugar-phosphate strand and from the overlap of $\pi$ electrons on adjacent bases along the stack. Such requirement is however fulfilled by the stacking potential in Eq.~(\ref{eq:01}) which has been
modified (with respect to the ladder Hamiltonian) so as to incorporate bending and twisting variables consistently with our geometrical model in Fig.~\ref{fig:1}. 
In particular,  it has been shown that the twist angle between adjacent bases introduces a restoring force in the stacking which stabilizes the helical molecule also in the presence of base pair fluctuations \cite{io12}. 

The robustness of our choice has also been tested by comparing model predictions and available experimental information, markedly looping probabilities and twist-stretch profiles in the presence of a constant load \cite{io16b,io18}.

Altogether, the potentials in  Eq.~(\ref{eq:01}) represent a realistic representation of the effective forces at play in DNA albeit in the context of a coarse-grained description which treats the nucleotide as a point-like object and does not account for some structural deformations such as the presence of grooves known to be important e.g., in the sequence specific DNA-protein binding \cite{bres07}.

Then, the Hamiltonian of our model is given by:

\begin{eqnarray}
& &H =\, H_a[r_1] + \sum_{i=2}^{N} H_b[r_i, r_{i-1}, \phi_i, \theta_i] \, , \nonumber
\\
& &H_a[r_1] =\, \frac{\mu}{2} \dot{r}_1^2 + V_{1}[r_1] \, , \nonumber
\\
& &H_b[r_i, r_{i-1}, \phi_i, \theta_i]= \,  \frac{\mu}{2} \dot{r}_i^2 + V_{1}[r_i] + V_{2}[ r_i, r_{i-1}, \phi_i, \theta_i]  \, , \nonumber
\\ 
\label{eq:01a}
\end{eqnarray}

where the term $H_a[r_1]$ is taken out of the sum as the first base pair in the chain lacks the preceding neighbor (but it interacts with its successive neighbor) along the molecule stack.

While this work focuses primarily on the global flexibility properties of short molecules as a function of length, the base pair specificities are not  considered in the following calculations. Accordingly we take a set of  model parameters, suitable to homogeneous sequences, in the range of those previously determined by fitting thermodynamic and elastic data \cite{krueg06,zdrav06,fenn08,weber13,singh15,weber15,io18}  i.e.,  $D_i=\,60 meV$, $b_i= 5 \AA^{-1}$,  $f_s=\,0.1$, $l_s=\,0.5 \AA$,  $K_S=\,10 mev \AA^{-2}$, $\rho_{i} \equiv \, \rho_{i, i-1} =\,1$, $\alpha_{i}\equiv \, \alpha_{i, i-1} =\, 2 \AA^{-1}$. 

The model can be extended to deal with the flexibility of DNA random sequences by introducing heterogeneity in the parameters of the one particle- and two particles potential analogously to what done in the study of bubble formation in specific minicircles \cite{io14a}.

\subsection*{C. \, Partition function}

The equilibrium properties for the system described by Eq.~(\ref{eq:01}) are investigated by path integral techniques \cite{io09,io10}. 
Essentially, the method relies on the assumption that the base pair displacement $r_i$ may be treated as an imaginary time dependent path $r_i(\tau)$ with $\tau=\,it$ and $t$ being the real time for the path evolution amplitude within a given time range \cite{fehi}. The mapping to the imaginary axis is a well-known semi-classical technique in the solution of quantum mechanical problems \cite{jack}: namely, one first searches for that classical path which minimizes the sum of the kinetic and potential energy in the Euclidean action and then, around such classical path, one evaluates the quantum fluctuation contribution to the path integral. 

Likewise, the space-time mapping technique has also been widely applied to derive the partition function $Z_N$ for the DNA molecules keeping in mind that the latter represent a classical system \cite{io11a}. Specifically, this has been done using the same method which permits solving the pseudo-Schr\"{o}dinger equation for a Morse potential obtained from the Hamiltonian in Eqs.~(\ref{eq:01}),~(\ref{eq:01a})  after suppressing angular variables, non-linear parameters and taking the strong $K_S$ limit \cite{landau,pey04,hand12,io14b}.

Inherent to the method is that $Z_N$ is written as an integral over closed paths, $r_i(0)=\,r_i(\beta )$, where $\beta$ is the inverse temperature which sets the amplitude of the trajectories along the $\tau$-axis. 
Accordingly any base pair trajectory $r_i(\tau)$ can be Fourier expanded in a series whose coefficients yield an ensemble of possible radial fluctuations which statistically contribute to the path integration with their specific Boltzmann weight, provided they fulfill the physical requirements set by the model potential and outlined above.  

Accordingly, integrating over the Fourier coefficients, the program adds more and more trajectories to $Z_N$ until numerical convergence is achieved. This defines the state of thermodynamic equilibrium for the system. As our model contains bending and twisting degrees of freedom, the convergence must be checked also against the angular variables with their specific cutoffs.

Then, the general $Z_N$ associated to Eq.~(\ref{eq:01a}) is given by:

\begin{eqnarray}
& &Z_N=\, \oint Dr_{1} \exp \bigl[- A_a[r_1] \bigr]   \prod_{i=2}^{N}  \int_{- \phi_{M} }^{\phi_{M} } d \phi_i \int_{- \theta_{M} }^{\theta _{M} } d \theta_{i} \oint Dr_{i}  \exp \bigl[- A_b [r_i, r_{i-1}, \phi_i, \theta_i] \bigr] \, , \nonumber
\\
& &A_a[r_1]= \,  \int_{0}^{\beta} d\tau H_a[r_1(\tau)] \, , \nonumber
\\
& &A_b[r_i, r_{i-1}, \phi_i, \theta_i]= \,  \int_{0}^{\beta} d\tau H_b[r_i(\tau), r_{i-1}(\tau), \phi_i, \theta_i] \, ,
\label{eq:02}
\end{eqnarray}

and the free energy of the system is computed from Eq.~(\ref{eq:02}) as: $F=\, -\beta ^{-1} \ln Z_N$. 

The cutoffs on the bending and twisting angles are set to $\phi_{M}=\,\pi /2$ and $\theta_{M}=\,\pi /4$, respectively. These choices are consistent with the formation of kinks which locally reduce the bending energy \cite{kim14,sung15} and permit achieving numerical convergence in the computation of Eq.~(\ref{eq:02}).
 
While $A_b$ is indeed the sum of the kinetic and potential energy, the integration measure $\oint {D}r_i$ over the space of the Fourier coefficients is chosen to normalize the kinetic action according to the condition \cite{io03,io05}: 

\begin{eqnarray}
\oint {D}r_i \exp\Bigl[- \int_0^\beta d\tau {\mu \over 2}\dot{r}_i(\tau)^2  \Bigr] = \,1 \, ,
\label{eq:03} \,
\end{eqnarray}

which also sets the free energy zero. 
It is emphasized that, as the normalization condition holds for any $\mu$, $F$ does not depend on $\mu$ as expected for a classical system.

Eqs.~(\ref{eq:02}),~(\ref{eq:03}) define the ensemble over which we carry out the statistical averages for the physical parameters and distribution functions of the DNA fragments.

From Eq.~(\ref{eq:03}), one derives the temperature dependent cutoffs in the integration over the radial fluctuations as detailed in ref.\cite{io11a}. In this regard two considerations follow:

\textit{i)} As the base pair displacements sample a temperature dependent portion of the phase space defined in Eq.~(\ref{eq:02}), the temperature effects on the physical properties of the molecules can be suitably incorporated in the path integral computation.
The calculations in Sections III and IV are performed at room temperature.

\textit{ii)} The cutoff on the radial base pair fluctuations, measured with respect to $R_0$, is set at $12$\AA. Hence, large fluctuations and base pair breaking are incorporated in the code.

Finally we notice that, as the path integration consistently truncates the phase space available to the base pair displacements, one may extend the method
to tackle problems in which the free volume for the DNA molecule is restricted by the presence of crowders as it occurs in cells \cite{chers15}. This may also be done by considering site specific effects of non uniformly distributed crowders on DNA oligomers.

\section*{III. \, Twist conformation}

As mentioned in the Introduction, the DNA biological functioning is intrinsically associated to the fluctuational opening of base pairs and formation of transient bubbles \cite{gueron,benham99,bonnet03,zocchi03,rapti06,kalos09,kalos11,singh11,io13} which locally untwist the helix leading to a state of negative supercoiling \cite{metz10}. DNA torsional and bending flexibility has been widely investigated by techniques which stretch single molecules and study their response as a function of the applied load \cite{busta92,marko97,busta06,mameren09,wuite11,marko15}. In all these cases, both in vivo and in vitro, the average helical repeat ($< h >$) may deviate from the value associated to the unperturbed configuration. It seems therefore convenient to have a theoretical scheme which treats the helical repeat ($h$) i.e., the number of base pairs per helix turn, as an input parameter associated to a specific helical conformation and determines \, $< h >$ \, by summing over the ensemble of all base pair fluctuations contributing to Eq.~(\ref{eq:02}). 

To this purpose,  we have devised a recursive method which defines the twist angle $\theta_i$ of the $i-th$ base pair (in Fig.~\ref{fig:1}) as the sum of the average $<\theta_{i - 1}>$ computed for the preceding base pair along the axis and the increment $2\pi / h$ accounting for the molecule helical conformation.  For the first base pair in the chain, the average twist angle is set equal to zero ($< \theta_1 >=\,0$).  Furthermore, $h$ is tuned within a broad range ($ \in [6, \,14]$) and, for any selected $h$, an integration is performed over a twist fluctuation $\theta_{i}^{fl}$ around the value $\, <\theta_{i - 1}>  + 2\pi / h \,$. Thus, $\theta_i$  in Eq.~(\ref{eq:02}) is written as, \, $\theta_i =\, <\theta_{i - 1}>  + 2\pi / h + \theta_{i}^{fl}$ \, and the cutoff $\theta_{M}$ refers to the integration over the twisting fluctuation variable $\theta_{i}^{fl}$ \cite{io17}. 

With this procedure, given a set of $n$ input values for $h$,  we derive a set of $n$ average helical repeat as:

\begin{eqnarray}
< h >_{j}=\,\frac{2\pi N}{< \theta_N >} \, , \, \hskip 1cm  (j=\,1,...,n) ,
\label{eq:04}
\end{eqnarray}

where $< \theta_N >$ is the average twist for the last base pair in the chain.
It is emphasized that the ensemble averaged twist angles also incorporate the contributions of the radial and bending fluctuations which are intertwined in the computation as made evident in Eq.~(\ref{eq:02}).  
The accuracy of the method has been tested by increasing the number $n$ of values sampled by the program until convergence is achieved for the free energy of the molecule conformation associated to any $< h >_{j}$. 

By free energy minimization, one eventually determines the equilibrium average helical repeat ($< h >_{j*}$)  in specific environmental conditions defined e.g. by the temperature, salt concentration and in the presence of external forces. Besides, one can study the thermodynamics of the molecule both in the untwisted ($< h >_{j}$  larger than  $< h >_{j*}$) and in the over-twisted ($ < h >_{j}$ smaller than  $< h >_{j*}$) conformations.

\begin{figure}
\includegraphics[height=8.0cm,width=8.0cm,angle=-90]{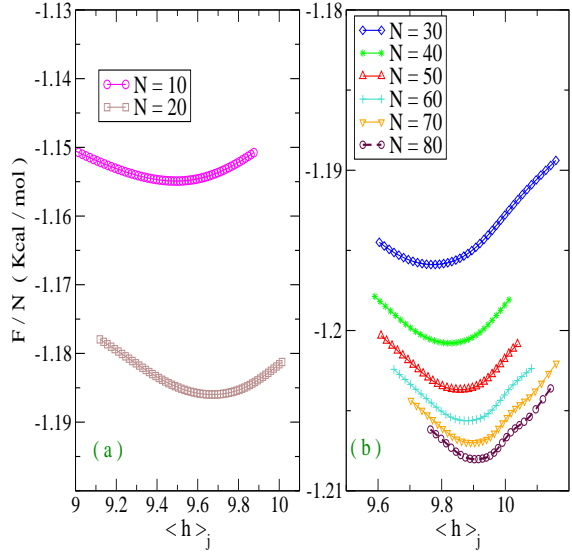}
\caption{\label{fig:2}(Color online)  
Free energy per base pair calculated, at room temperature, as a function of the average helical repeat for the molecules with (a) {} 10 and 20 base pairs; (b) {} 30 to 80 base pairs.
}
\end{figure}

\begin{figure}
\includegraphics[height=8.0cm,width=8.0cm,angle=-90]{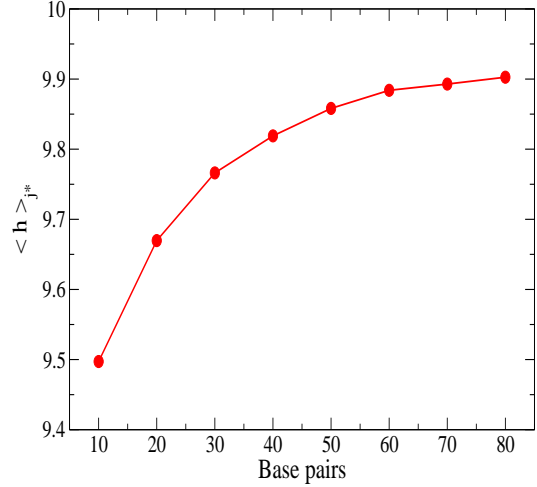}
\caption{\label{fig:3}(Color online) 
Equilibrium average helical repeat, obtained by free energy minimization, for the molecules in Fig.~\ref{fig:2}.
}
\end{figure}

In Fig.~\ref{fig:2}, we plot the free energy per base pair against $< h >_{j}$ for a set of eight homogeneous fragments with $N =\,10, ..., 80$. It appears that, by increasing $N$, the $F / N$ values generally decrease for any twist conformation indicating that the molecule stability grows with the length. Moreover, for larger $N$, the  plots also get more densely spaced pointing to a convergence in the free energy values per base pair for molecules with $\sim 10^{2}$  base pairs. Noticeably, the $F / N$  minima which determine the $< h >_{j*}$'s shift upwards by increasing the molecule length. As highlighted in Fig.~\ref{fig:3}, the $< h >_{j*}$'s grow versus $N$ with large gradient for very short sequences and with a slower pace for longer fragments. Although the computations has not been carried out for $N > 80$, $< h >_{j*}$ is expected to converge to $\sim 10$ for molecules 
of order $\sim 10^{2}$  base pairs, thus approaching the standard room temperature experimental value of kilo base long DNA \cite{wang79,duguet93}.

Clearly the specific twist conformation affects the molecule structural parameters such as average diameter and elongation \cite{bohr11}. Thus, the program calculates, for each $ < h >_{j}$, also the ensemble averages for the base pair distances in Eq.~(\ref{eq:0}) \, ( $< \overline{d_{i,i-1}} >$ ) \, and for the bending angles ( $< \phi_{i} >$ ) which are essential to the analysis of the contour length and radial distribution functions. 
It is understood that the ensemble averages are carried out according to the procedure described in Section II and summarized by Eqs.~(\ref{eq:02}),~(\ref{eq:03}).

In the following we assume that the DNA fragments are in the equilibrium twist conformation determined by $< h >_{j*}$ and, accordingly, calculate the distribution functions using the ensemble averaged helical parameters associated to such conformation.

\section*{IV. Distribution Functions}

\subsection*{A. Contour length}

The contour length distribution function $C(L)$ is defined as:

\begin{eqnarray}
& &C(L)=\,  \biggl<  \delta \biggl( \sum_{i=2}^{N} \bigl | \overline{d_{i,i-1}} \bigr | - L \biggr) \biggr > 
\label{eq:05}
\end{eqnarray}

where $\delta (x - L)$ is the Dirac $\delta$-function.

In order to compute $C(L)$, we first build a physically meaningful range for the contour length  $L$: for any dimer, fluctuations of variable amplitude $\Delta_a$ are considered with respect to the ensemble averaged $< \overline{d_{i,i-1}} >$ computed as described in Section III. Then, after setting \, $\bigl |\overline{d_{i,i-1}} \bigr | =\, < \overline{d_{i,i-1}} > \pm  \Delta_a$, \,  $L$  in Eq.~(\ref{eq:0a}) can be written as:

\begin{eqnarray} 
& &L=\, < L >_{N} \pm  ( N - 1) \Delta_a \, , \nonumber
\\ 
& &< L >_{N}=\, \sum_{i=2}^{N} < \overline{d_{i,i-1}} > \, , 
\label{eq:06}
\end{eqnarray}

where $< L >_{N}$ is the mean contour length for the fragment with $N$ base pairs. 
Then, using the integral representation for the Dirac $\delta$-function, Eq.~(\ref{eq:05}) transforms into:

\begin{eqnarray}
& &C(L^{ })=\,  \biggl< (2\pi )^{-1} \int _{-\infty}^{\infty} dt  \exp \biggl[ i \sum_{i=2}^{N} \biggl( \bigl |\overline{d_{i,i-1}}\bigr | - < \overline{d_{i,i-1}} > \pm  \Delta_a \biggr) t \biggr] \biggr> \, , 
\label{eq:07}
\end{eqnarray}

which is computed by taking again the averages over the ensemble specified by Eq.~(\ref{eq:02}).  Note that the $\exp$-argument can be written in the form of a sum of decoupled contributions from successive dimers as $< L >_{N}$ itself is a sum over the base pair index of ensemble averaged terms.

\subsection*{B. End-to-End Distribution Function}

For a given value of $L$, the radial end-to-end distribution function  $G(R_{e-e},\, L)$ is defined by:

\begin{eqnarray}
& &4 \pi R^2_{e-e} G(R_{e-e},\, L)=\, \biggl < \delta \biggl( \biggl| \sum_{i=2}^{N}  \overline{d_{i,i-1}} \biggr| - R_{e-e} \biggr) \biggr > \, . 
\label{eq:08}
\end{eqnarray}

Similarly to what has been done for the contour length, we need now to set a suitable range of $R_{e-e}$'s values around its ensemble average for the calculation of $G(R_{e-e},\, L)$. However, one notices that the end-to-end distance in Eq.~(\ref{eq:0a}) and its ensemble average cannot be decoupled as sums over the base pair index. 
Hence, the procedure used to derive Eq.~(\ref{eq:07}) cannot be applied to transform Eq.~(\ref{eq:08}) in a form suitable for computation.

The problem is circumvented by observing that, for short sequences  whose molecular axis lies on a plane \cite{n3},  $R_{e-e}^{2}$ is indeed  approximated by a sum over $i$ as double index terms of order $\cos(\phi_{j}-\phi_{i})$, yielding minor contributions to the statistical averages over bending fluctuations, can be neglected. Explicitly, from the second of Eqs.~(\ref{eq:0a}), we find that:

\begin{eqnarray} 
& &R_{e-e}^{2} \equiv \biggl| \sum_{i=2}^{N}  \overline{d_{i,i-1}} \biggr|^2 = \,  \sum_{i=2}^{N} R_{e-e}(i)^{2}  \, , \nonumber
\\ 
& &R_{e-e}(i) \simeq \sqrt{ d_{i,i-1}^2 + 2 d_{2,1}d_{i,i-1} \cos(\phi_{i-1}) } \, .
\label{eq:08b}
\end{eqnarray}

Besides decoupling the single dimer contributions, Eq.~(\ref{eq:08b}) also defines a quantity,  $R_{e-e}(i)$,  which is linear in the distances $d_{i,i-1}$. Then,  for any dimer, we take fluctuations of amplitude $\Delta_b$ with respect to its average \, $< R_{e-e}(i) >$, i.e., \, $R_{e-e}(i)=\, < R_{e-e}(i) > \pm  \Delta_b$ \, and set:

\begin{eqnarray} 
& &< R_{e-e} >_{N} = \, \sqrt{ \sum_{i=2}^{N} \bigl <  R_{e-e}(i)  \bigr >^{2} } \, .
\label{eq:08c}
\end{eqnarray}

Note that Eq.~(\ref{eq:08c}) may differ from the exact mean value which, in principle, can be obtained by ensemble averaging the second of Eqs.~(\ref{eq:0a}). 
This however is not relevant to our purposes, as Eq.~(\ref{eq:08b}) and Eq.~(\ref{eq:08c}) consistently provide a reference value around which, by varying $\Delta_b$, we can sample a range of $R_{e-e}$'s for the calculation of  $G(R_{e-e},\, L)$.

Then, applying  the identity:
\begin{eqnarray} 
\delta (x^2 - R_{e-e}^2) = \frac{1}{2 R_{e-e}} \bigl[ \delta (x + R_{e-e}) + \delta (x - R_{e-e})\bigr] \, ,
\label{eq:09}
\end{eqnarray}
with $x \equiv \biggl| \sum_{i=2}^{N}  \overline{d_{i,i-1}} \biggr|$,  we first re-write Eq.~(\ref{eq:08}) as
\begin{eqnarray}
& &4 \pi R^2_{e-e} G(R_{e-e},\, L)=\, 2 R_{e-e} \biggl < \delta \biggl( \biggl| \sum_{i=2}^{N}  \overline{d_{i,i-1}} \biggr|^2 - R_{e-e}^2 \biggr) \biggr > \, . 
\label{eq:08d}
\end{eqnarray}

Finally, using the integral representation for the Dirac $\delta$-function,  Eq.~(\ref{eq:08d}) is transformed in the form: 

\begin{eqnarray}
& &4 \pi R^2_{e-e}G(R_{e-e}^{ },\, L) = \,  \biggl< \frac{R_{e-e}}{\pi } \int _{-\infty}^{\infty} du  \exp \biggl[ i \biggl(  \sum_{i=2}^{N} \biggl( R_{e-e}(i)^{2}  -  [ < R_{e-e}(i) > \pm  \Delta_b ]^{2}  \biggr) \biggr) u \biggr] \biggr> \, \, . \nonumber
\\ 
\label{eq:010}
\end{eqnarray}

Hereafter Eq.~(\ref{eq:010}) is computed with the above defined ensemble averages and assuming for the contour length the value $< L >_{N}$ in Eq.~(\ref{eq:06}).

It is also noticed that, by integrating over $L$ the product of Eq.~(\ref{eq:05}) and Eq.~(\ref{eq:08}), one can numerically derive the full radial distribution function $G(R_{e-e}^{ })$. This step however, which would further increase the computational time, is beyond the scope of this work. On the other hand, the most probable end-to-end distance calculated by Eq.~(\ref{eq:010}) is not expected to deviate substantially from the maximum value of $G(R_{e-e}^{ })$.

\subsection*{C. Results}

\begin{figure}
\includegraphics[height=8.0cm,width=8.0cm,angle=-90]{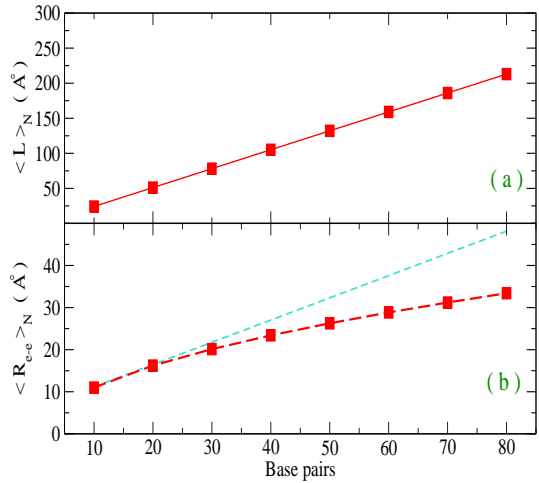}
\caption{\label{fig:4}(Color online) (a) Mean contour length calculated by Eq.~(\ref{eq:06}) and (b) mean end-to-end distance calculated by Eq.~(\ref{eq:08c}), for the set of eight fragments studied in Fig.~\ref{fig:2}. The dashed line is a guide to the eye to emphasize how $< R_{e-e} >_{N}$ deviates from linearity versus the fragment length.
}
\end{figure}

In Fig.~\ref{fig:4}, we plot the mean contour length and end-to-end distance, calculated by Eq.~(\ref{eq:06}) and  Eq.~(\ref{eq:08c}) respectively, for the short sequences considered in Section II.
While $< L >_{N}$ is obviously a linear function of $N$, $< R_{e-e} >_{N}$ significantly deviates from linearity as the molecules become longer. This follows from the fact that coiled configurations are entropically favored and their statistical weight in the partition function is larger in longer chains as expected. 
Such physical interpretation, extracted by
Fig.~\ref{fig:4}(b), indicates that Eq.~(\ref{eq:08c}) is a good approximation of the mean end-to-end distance and that the assumptions made to derive Eq.~(\ref{eq:08b}) are reliable.

\begin{figure}
\includegraphics[height=8.0cm,width=8.0cm,angle=-90]{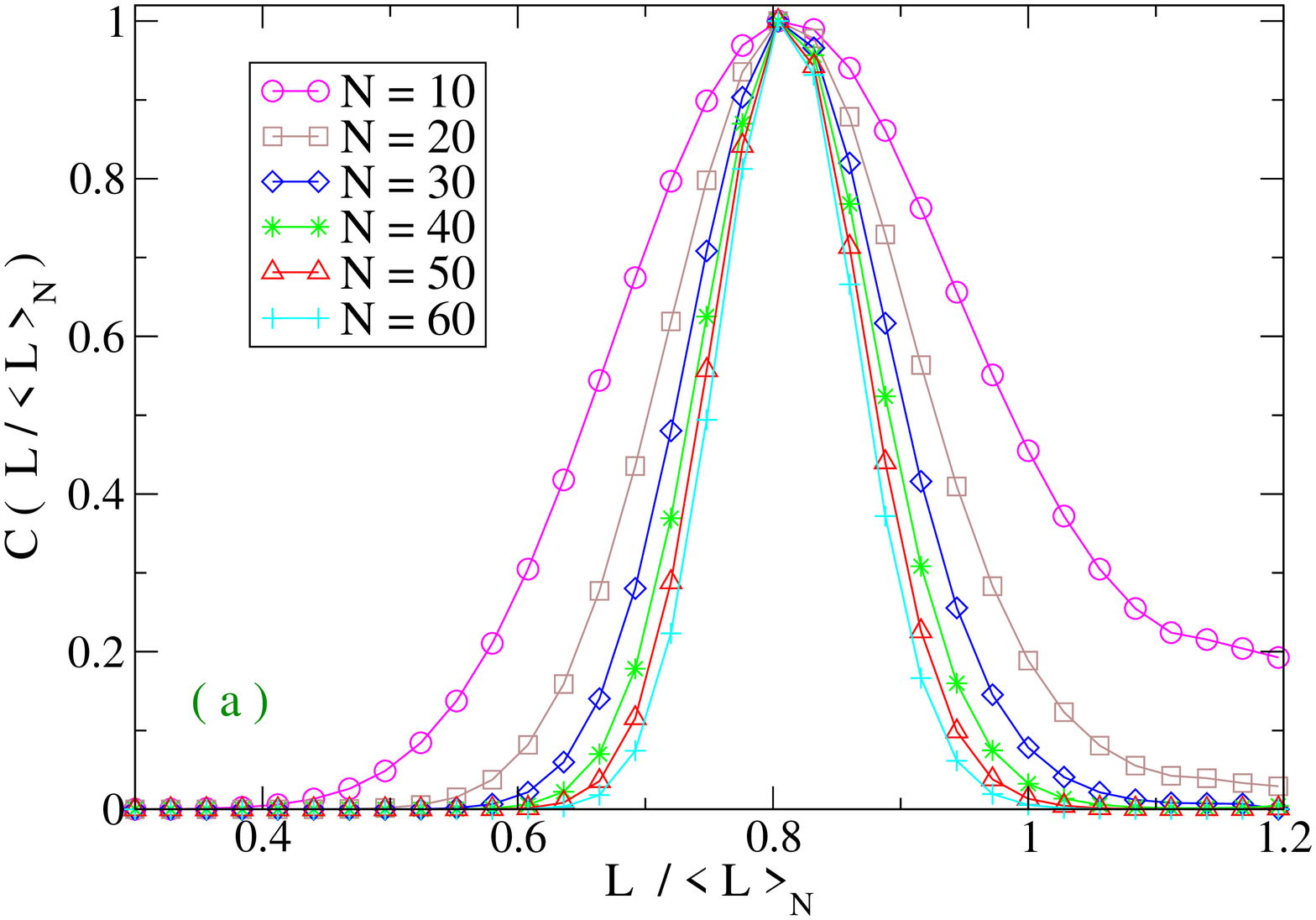}
\includegraphics[height=8.0cm,width=8.0cm,angle=-90]{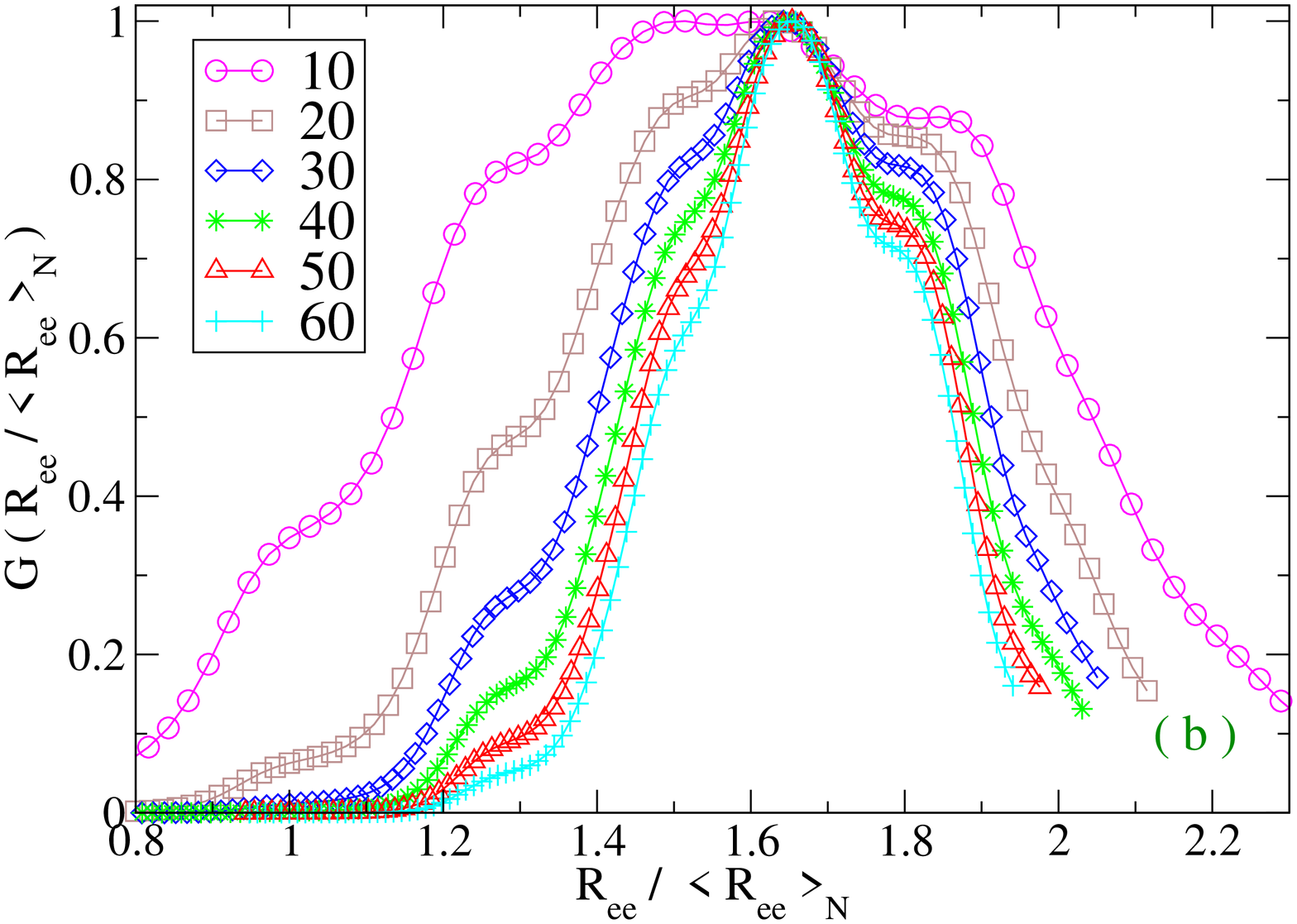}
\caption{\label{fig:5}(Color online)  
(a) {} Contour length distribution function from Eq.~(\ref{eq:07}) and (b) {} radial distribution function from Eq.~(\ref{eq:010}), for six short fragments. For each curve, the values are normalized to its peak value.
}
\end{figure}

In Fig.~\ref{fig:5}, we plot the contour length and the radial end-to-end distance distribution functions, normalized to the peak value, for a set of six short sequences. The contour length abscissa is $L /< L >_{N}$ while the end-to-end distance abscissa is $R_{e-e} /< R_{e-e} >_{N}$ to highlight the evolution in the shape of the distribution functions versus $N$.
Both functions approach a Gaussian statistics by increasing $N$ although the convergence of the contour length distribution to a bell-shaped curve is smoother: in fact it is the end-to-end distance to be mainly affected by fluctuational effects which cause the presence of bumps, mainly for very short chains. 

For $N \sim 50$ and beyond, the distributions tend to overlap. Note instead that in the simple random chain model for a finite chain with $N$ joints, the end-to-end distribution shows fast convergence, already for $N \sim 10$, to the Gaussian distribution obtained in the continuum limit \cite{n4}.

\begin{figure}
\includegraphics[height=8.0cm,width=8.0cm,angle=-90]{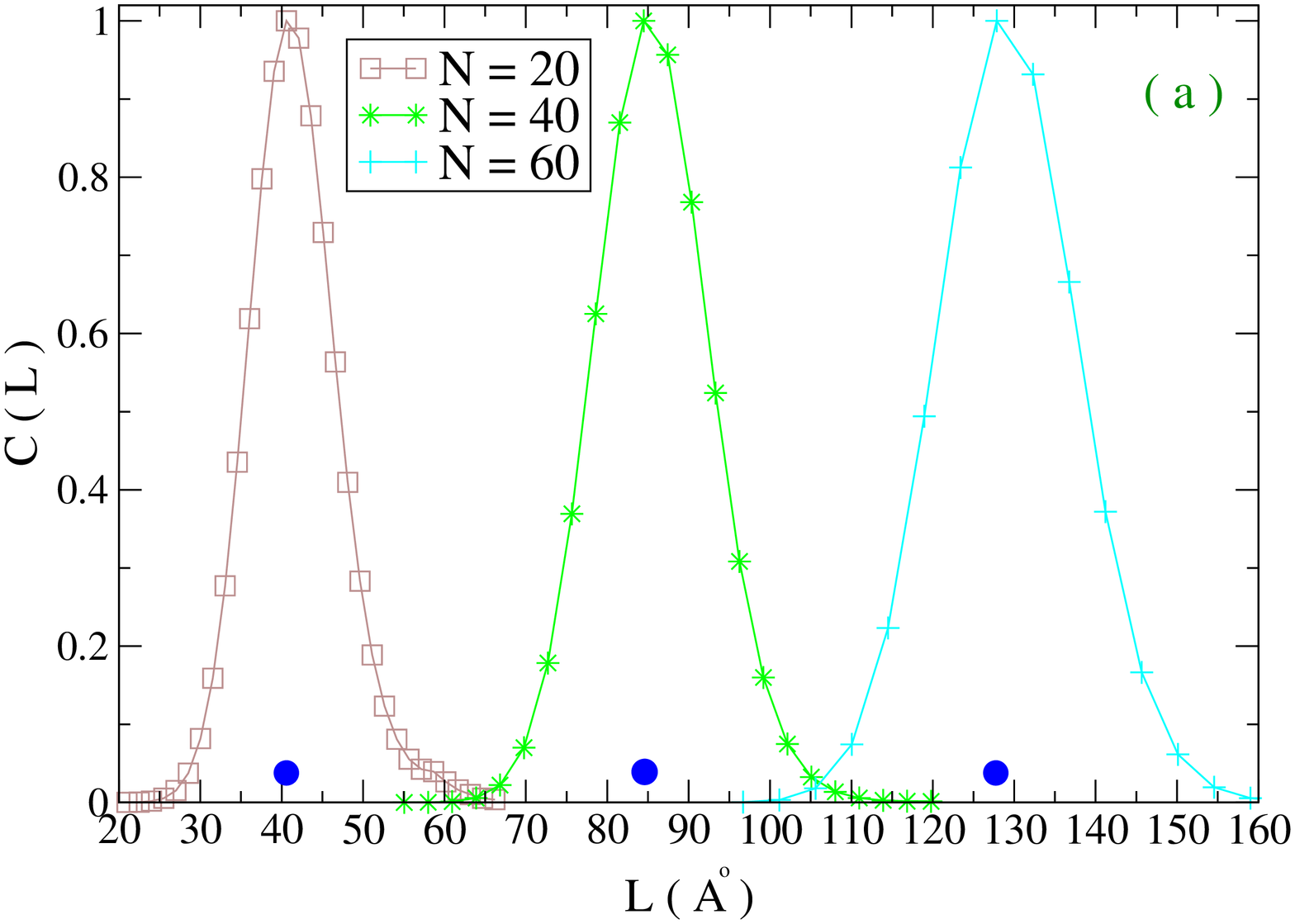}
\includegraphics[height=8.0cm,width=8.0cm,angle=-90]{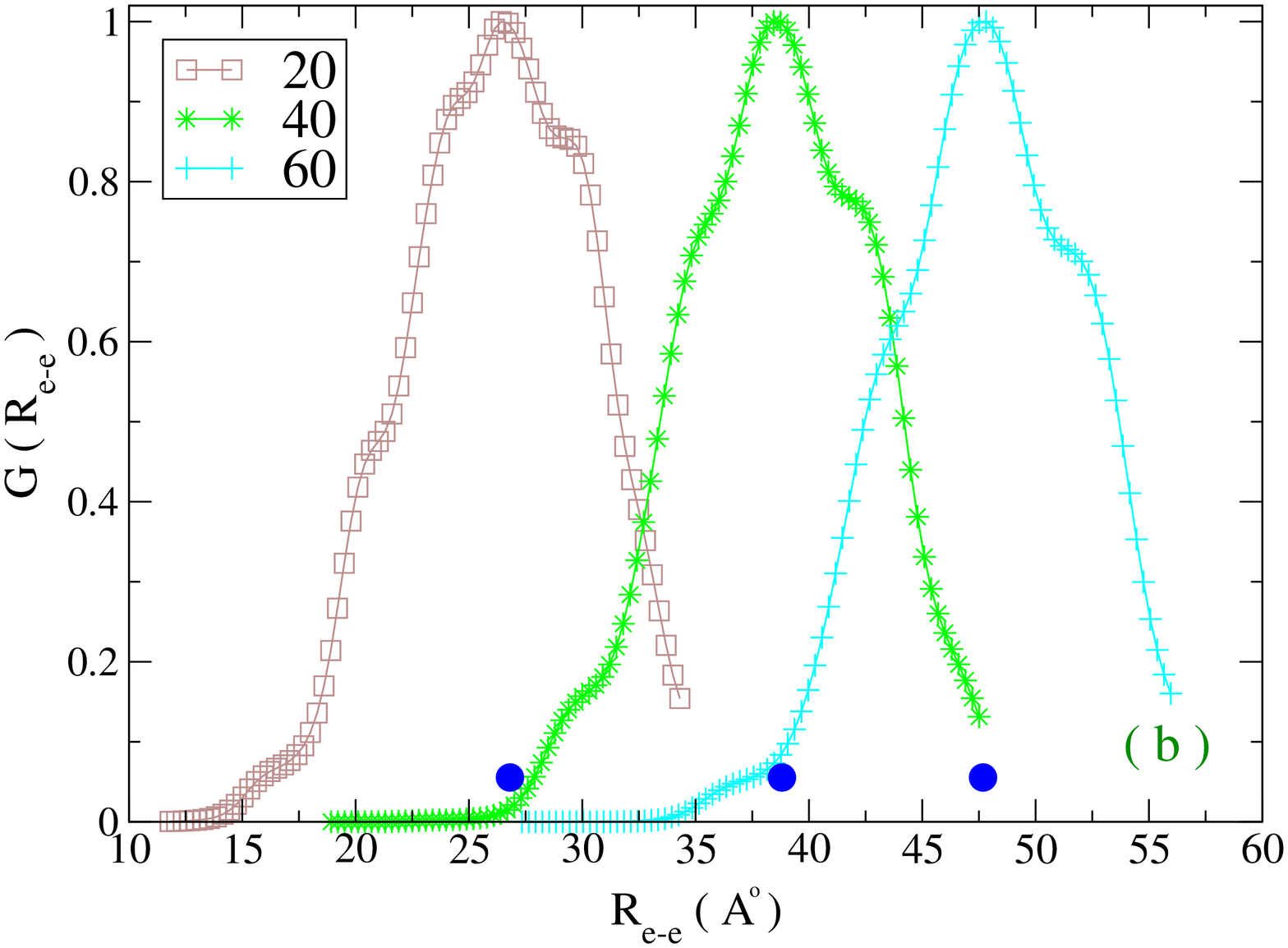}
\caption{\label{fig:6}(Color online)  
(a) {} Contour length distribution function versus $L$ and (b) {} radial distribution function versus $R_{e-e}$, for three short fragments. The blue dots mark the most probable value for each distribution. For each curve, the values are normalized to its peak value.
}
\end{figure}

It is useful to display the distribution functions as a function of $L$ and $R_{e-e}$ respectively, regardless of the  $< L >_{N}$'s and  $< R_{e-e} >_{N}$'s values, as done in Fig.~\ref{fig:6}. Only three sequences are considered for clarity. The blue dots mark in both panels the most probable value i.e., the maximum of the distribution functions for each sequence. While the maximum values  of the contour length distributions ($L^{M}$) scale linearly with $N$, the maximum values  of the end-to-end distributions ($R_{e-e}^{M}$) get more closely spaced by increasing $N$.   This confirms the indication put forward by Fig.~\ref{fig:4}(b).

In particular, for $N=\,20$, we get $L^{M}=\,40.58$\AA \, and $R_{e-e}^{M}=\, 26.43$\AA; for $N=\,40$,  $L^{M}=\,84.48$\AA \, and $R_{e-e}^{M}=\,38.44$\AA; for $N=\,60$,  $L^{M}=\,127.86$\AA \, and $R_{e-e}^{M}=\,47.81$\AA.  Hence, the ratio $R_{e-e}^{M} / L^{M}$ decreases as a function of $N$ signaling that longer molecules are intrinsically more flexible and can be bent more easily. Note however that, even for $N=\,20$,  we find \, $R_{e-e}^{M} / L^{M} < 1$  indicating that for such short fragments a fully stretched conformation is unlikely to occur. In fact, the physical source of this result lies in the fact that the molecular bonds are constantly bent under the effect of the thermal fluctuations and this feature is captured by our computational method.

Providing an estimate of the coil size, the computed $R_{e-e}^{M}$'s and $L^{M}$'s values in Fig.~\ref{fig:6} may be used to extract information on the persistence length ($l_p$) of the respective chains. This can be done e.g., in the framework of the WLC model, widely used to fit experimental data, in which the persistence length defines the characteristic length over which the infinitesimally small chain bonds maintain an exponentially decaying orientational correlation. Accordingly, keeping the contour length $L$ finite  while $N$ gets infinitely large, the WLC end-to-end distance reads:

\begin{eqnarray}
R_{e-e}^2 =\,  \int_{0}^{L}ds \int_{0}^{L}ds' \exp \biggl(-\frac{|s - s'|}{l_p} \biggr) \, ,
\label{eq:011}
\end{eqnarray}

where $s$, $s'$ are the arc length variables along the chain  \cite{kp}. 

As $R_{e-e}$ represents a global conformational property of the chain, $l_p$ defined by Eq.~(\ref{eq:011}) incorporates contributions from both short and long range interactions hence, it depends on $L$  which, in the WLC model, is usually assumed much larger than $l_p$ \cite{croq99}. 
Although it is conceptually distinct from the \textit{true} persistence length which measures strictly local correlations \cite{odi,skofix} (and should not depend on $L$),  the apparent $l_p$ in Eq.~(\ref{eq:011}) is directly related to the observable $R_{e-e}$ and therefore it can be used for our purpose. For more detailed analysis of the persistence length in flexible polymers with bending fluctuations and in charged polymers with variable temperature, ionic strength and salinity, we refer to \cite{reed91,thirum95,soder97,manni06,cifra10,volo11,save12,mazur14,manghi15}. 

Incidentally it is also observed that,  by tuning the parameters of the solvent term in 
Eq.~(\ref{eq:01}), one may study the flexibility properties as a function of the salt concentration  similarly to what done in ref.\cite{chers11} for the cyclization efficiency.

\begin{figure}
\includegraphics[height=8.0cm,width=8.0cm,angle=-90]{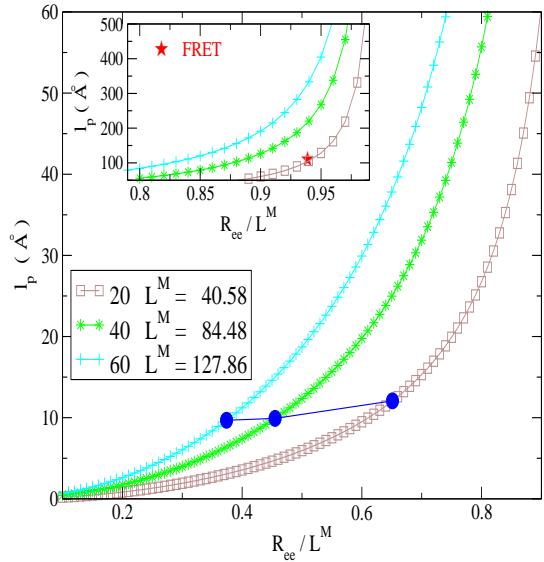}
\caption{\label{fig:7}(Color online)  Apparent persistence length as a function of the end-to-end distance $R_{e-e}$ calculated by Eq.~(\ref{eq:010}) for the three chains in Fig.~\ref{fig:6}.  For each chain, $L$ is set equal to the most probable contour length, $L^{M}$, marked by the blue dot in Fig.~\ref{fig:6}(a). Each $l_p$ value is determined with an accuracy $\leq 10^{-5}$. The blue dots on the curves correspond to the most probable end-to-end distances, $R_{e-e}^{M}$, shown by the blue dots in Fig.~\ref{fig:6}(b). The inset magnifies the upper range of $R_{e-e}$'s in which the chains tend to assume a straight rod conformation. Accordingly, in such range, the $l_p$'s get even larger than their respective $L^{M}$'s. The red symbol $\bigstar$ in the inset marks the $l_p$ value estimated in \cite{archer08} by fitting the FRET data for a set of fragments with $N \in [15,21]$.
}
\end{figure}

Eq.~(\ref{eq:011}) can be numerically solved to derive $l_p$ as a function of the ratio  $R_{e-e} / L$ which is a measure of the chain bending flexibility. We consider 
the three chains of Fig.~\ref{fig:6} and take the respective most probable lengths $L=\,L^{M}$ shown in Fig.~\ref{fig:6}(a). As displayed in Fig.~\ref{fig:7},  three plots are obtained for $l_p$ against the ratio $R_{e-e} / L^M$, with variable $R_{e-e}$.  Once $R_{e-e}$ is pinned to the $R_{e-e}^{M}$'s shown in Fig.~\ref{fig:6}(b), one gets the $l_p$ values for the three chains marked by the respective blue dots. Analogous trend and only slightly different $l_p$'s would be obtained by taking $L$ equal to the mean contour lengths in Fig.~\ref{fig:4}.

From Fig.~\ref{fig:7}, some considerations ensue:

i) $l_p$  grows versus the ratio $R_{e-e} / L^M$ consistently with the fact that straighter chains are stiffer. 
However, for a given value of $R_{e-e} / L^M$, $l_p$ is larger in longer chains. Then, the stiffness of the molecule (per base pair) should be rather measured by the ratio $l_p / L^{M}$.

ii) Generally, our estimated $l_p$'s  for short fragments are much smaller than the standard $l_p$ of kilo-base long DNA.

Precisely, for the shorter (and stiffer) chain in Fig.~\ref{fig:7} with $R_{e-e}^{M} / L^{M} \sim 0.65$, we find \, $l_p \sim 12$\AA. By increasing $N$, the calculated ratios $R_{e-e}^{M} / L^{M}$'s get smaller but this entropic effect is offset by the longer $L^{M}$'s hence,  $l_p$ remains of order $\sim 10$\AA. This holds for the short sequences here considered.  For $N=\,40$, we find \, $R_{e-e}^{M} / L^{M} \sim 0.46$.

iii) These results suggest some comparison with Ref.\cite{tan17} and specifically with its Fig.4: the plots are obtained on the base of the WLC model for a chain with $N=\,50$. The most probable $R_{e-e}$ is shown to vary with the persistence length  and, for $l_p=\,20$\AA, it is located at $R_{e-e}^{M} / \overline{L} \sim 0.45$, with $\overline{L}$ being the mean contour length which is equal, in Ref.\cite{tan17}, to the most probable contour length.
Notwithstanding that those plots  refer to the full end-to-end distribution (whereas our Eq.~(\ref{eq:010}) is calculated for a specific contour length i.e. the mean value), 
it appears that the very short $l_p$'s in Ref.\cite{tan17} are associated to coiled configurations with small ratios $R_{e-e}^{M} / L^{M}$, consistently with our results. It is however remarked that, in the simple WLC model of Ref.\cite{tan17}, $l_p$ (and the stretching modulus) is taken as an input parameter whereas in our analysis $l_p$ is an output of the computational method as it is extracted from the most probable values of the computed distribution functions.

The inset in Fig.~\ref{fig:7} magnifies the range $R_{e-e} / L^{M}$ corresponding to almost straight chains in which $l_p$ approaches the standard value, $500$\AA, of kilo-base DNA.  

Even larger $l_p$'s can be found by solving Eq.~(\ref{eq:011}) assuming the ratio $R_{e-e} / L^M =\,0.99$. In this limit, e.g. for $N=\,20$, Eq.~(\ref{eq:011}) yields  $l_p=\,669$\AA. 

Also the experimental \, $l_p^{\bigstar}=\,110$\AA \, \, derived by FRET measurements \cite{archer08} is reported on. As such value is obtained by fitting the variance of a Gaussian end-to-end distribution for a set of short chains with $N  \in [15,21]$, it seems appropriate to compare it with our plot for the chain $N=\,20$. This justifies the position of the red ${\bigstar}$ in the inset. 

While $l_p^{\bigstar}$ is substantially lower than the standard $500$\AA, it is still one order of magnitude larger than our above estimated $l_p$. The source of this discrepancy may be twofold.
On one side, the use of the Gaussian statistics to derive the relation between variance and contour length, which yields $l_p^{\bigstar}$,  may be questionable for very short fragments \cite{archer06}. For the latter,  as shown in Fig.~\ref{fig:5}, the end-to-end distribution is poorly approximated by a Gaussian law which instead is consistently recovered when the contour length is larger than $l_p$. On the other side, our estimate of $l_p$ is based on a relation, Eq.~(\ref{eq:011}), which pertains to a continuum model
whereas the most probable end-to-end distance and contour length have been computed by a discrete Hamiltonian model for an helix in a specific twist conformation. While this procedure has been followed in order to make some comparison with available estimates based on the WLC model, in a more precise and consistent analysis  $l_p$ should be directly computed by the same discrete and twisted model. This programme is left for a future work. 

Altogether the results here presented indicate that: (a) short DNA molecules are intrinsically flexible by virtue of strong radial and angular fluctuational effects, (b) assuming the WLC relation in Eq.~(\ref{eq:011}), $l_p$ can be estimated through the ratio $R_{e-e}^{M} / L^{M}$ where the most probable values are obtained respectively by the end-to-end  and by the contour length distribution functions.  
(c) for short molecules, $l_p$ may be at least one order of magnitude smaller than the standard value of kilo-base DNA.

\section*{VI. Conclusions}

This work investigates the DNA flexibility properties focusing on the contour length and radial distribution functions at those short length scales which are biologically important. In this range, we feel appropriate to use a method based on a mesoscopic Hamiltonian model which provides a realistic description of the helical structure at the level of the  base pair thus allowing for a broad range of possible radial and angular fluctuations.   The stacking potential describes the intra-strand forces between adjacent base pairs which stabilize the helix with the variable base pair distances depending both on the radial and angular variables. Accordingly, the molecules can be stretched, twisted and bent
without the need for those phenomenological parameters which account for the global flexibility in worm-like-chain models. Certainly our Hamiltonian depends on a set of  base pair effective parameters whose range of values, however, has been widely tested in previous works by fitting thermodynamic and elastic data. Although this work is restricted to homogeneous chains, the extension to incorporate sequence specificities in the model is straightforward.

The statistical method builds an ensemble of molecule conformations consistent with the model potential and achieves numerical convergence on the partition function  by including  $\sim 10^8$  configurations for each dimer along the molecule stack. Thus we are able to determine the molecule twist conformations with their specific free energy.

The computation has been carried out in two steps: first, for the equilibrium twist conformation associated to the free energy minimum, one obtains the ensemble averages for the structural parameters which determine the specific shape of the helix i.e., the base pair distances and the bending angles for each dimer in the sequence. Second,  one considers  all fluctuations around the mean values (for the contour length and the end-to-end distance) and performs the ensemble averages which contribute to the distribution functions in Eqs.~(\ref{eq:07}),~(\ref{eq:010}). In this way we have derived the distribution functions for a set of short molecules in the physically significant range defined by the average parameters and found, for each molecule, the most probable values for the contour length and for the end-to-end distance. These values and their ratio, being an indicator of the global flexibility of the chain,  have been used to examine the apparent persistence length as a function of the molecule size, also in relation with some experimental and numerical studies. Our results support the view that short molecules of $\sim 60$ base pairs (and less) maintain a remarkable flexibility by virtue of the statistical weight of the bent configurations and, accordingly, can have persistence lengths much smaller than those generally attributed to kilo-base long DNA.

While foremost task of this study has been that of developing a computational scheme for the distribution functions of short molecules, we finally mention that the method may be further applied to a number of cases which can be  experimentally accessed. For instance, one may study the evolution of the radial distribution function in the presence of an external load and/or under the effect of variable environmental conditions. The method may be also extended to molecules with $\sim 100$ base pairs and more, albeit at the price of a significant increase of the computational resources.


\begin{thebibliography}{widest-label}

\bibitem{hart10}
B. Heddi, J. Abi-Ghanem, M. Lavigne, B. Hartmann, \emph{J. Mol. Biol.}  \textbf{395}, 123-133 (2010).

\bibitem{albu14}
E.L. Albuquerque, U.L. Fulco, V.N. Freire, E.W.S. Caetano, M.L. Lyra, and F.A.B.F. de Moura,  \textit{Phys. Rep.} \textbf{535}, 
139 (2014).

\bibitem{noy16}
A. Noy, T. Sutthibutpong, S.A. Harris, \textit{Biophys. Rev.} \textbf{8}, 233-243 (2016).


\bibitem{dekker12}
I. De Vlaminck,  M.T.J. van Loenhout,  L. Zweifel,  J. den Blanken,  K. Hooning,  S. Hage, 
J. Kerssemakers,  C. Dekker, \textit{Molecular Cell} \textbf{46}, 616-624 (2012).


\bibitem{schleif92}
R. Schleif, \textit{Annu. Rev. Biochem.}  \textbf{61}, 199-223 (1992).

\bibitem{rich97} 
K. Luger, A.W. M\"{a}der, R.K. Richmond, D.F. Sargent, T.J. Richmond, \textit{Nature} \textbf{389}, 251-260 (1997).

\bibitem{prun98}
A. Prunell, \textit{Biophys. J.} \textbf{74}, 2531-2544 (1998).

\bibitem{wang12}
J.L. Killian, M. Li, M.Y. Sheinin, M.D. Wang,   \textit{Curr. Opin. Struct. Biol.}  \textbf{22}, 80–87 (2012).

\bibitem{licht14}
E.C. Small, L. Xi, J.-P. Wang, J. Widom, J.D. Licht, \emph{Proc. Natl. Acad. Sci. USA}  \textbf{111}, E2462-E2471  (2014).

\bibitem{croth90}
D.M. Crothers, T.E. Haran, J.G. Nadeau, \textit{J. Biol. Chem.} \textbf{265}, 7093-7096 (1990).

\bibitem{stell13}
E. Stellwagen, J.P. Peters, L.J. Maher, N.C. Stellwagen, \textit{Biochemistry} \textbf{52}, 4138-4148 (2013).

\bibitem{ohy01}
T. Ohyama, \textit{Bioessays} \textbf{23}, 708-715 (2001).

\bibitem{wigg07}
H.G. Garcia, P. Grayson, L. Han, M. Inamdar, J. Kondev, P.C.
Nelson, R. Phillips, J. Widom, P.A. Wiggins, \textit{Biopolymers}  \textbf{85}, 115-130 (2007).

\bibitem{chen10}
B. Chen, Y. Xiao, C. Liu, C. Li, F. Leng, \emph{Nucl. Acids Res.}  \textbf{ 38}, 3643-3654 (2010).


\bibitem{shore}
D. Shore, J. Langowski, R.L. Baldwin,   \emph{Proc. Natl. Acad. Sci. USA}  \textbf{78},  4833-4837   (1981).

\bibitem{horo}
D.S. Horowitz, J.C. Wang, \emph{J. Mol. Biol.} \textbf{173}, 75-91  (1984).

\bibitem{shimada}
J. Shimada, H. Yamakawa,   \emph{Macromolecules}  \textbf{ 17}, 689-698  (1984).

\bibitem{n2}
Altogether the $J$-factor is a non-monotonous function of the molecule length with the maximum located at $N \sim 500$ base pairs. Above such length scale, the $J$-factor smoothly decreases versus $N$ in accordance with the fact that open end conformations are progressively favored in longer molecules by entropic effects.

\bibitem{shore1}
D. Shore, R.L. Baldwin,   \emph{J. Mol. Biol.} \textbf{170},  957-981  (1983); ibid.,   \textbf{170}, 983-1007 (1983).


\bibitem{widom}
T.E. Cloutier, J. Widom,   \emph{Mol. Cell.}  \textbf{ 14}, 355-362 (2004).

\bibitem{vafa}
R. Vafabakhsh, T. Ha,     \emph{Science}  \textbf{337}, 1097-1101 (2012).

\bibitem{kim13}
T.T. Le, H.D. Kim, \emph{Biophys. J.}  \textbf{104},  2068-2076   (2013).

\bibitem{archer08} 
C. Yuan,  H. Chen, X.W. Lou,  L.A. Archer, \textit{Phys. Rev. Lett.} {\bf 100}, 018102 (2008).



\bibitem{gole12} 
A. Noy, R. Golestanian,   \textit{Phys. Rev. Lett.}   {\bf 109}, 228101 (2012).


\bibitem{maiti15}
A. Garai,  S. Saurabh,  Y. Lansac,  P.K. Maiti, \textit{J. Phys. Chem. B}, \textbf{119}, 11146-11156, (2015).

\bibitem{io11}
M. Zoli,       \emph{J. Chem. Phys.}  \textbf{135},  115101 (2011).


\bibitem{io14}
M. Zoli, \textit{J. Chem. Phys. } {\bf 141}, 174112 (2014).

\bibitem{io16b}
M. Zoli,   \textit{J. Chem. Phys.}   {\bf 144},  214104 (2016). 

\bibitem{volo08}
Q. Du,  A. Kotlyar, A. Vologodskii,   \emph{Nucl. Acids Res.}  \textbf{ 36},  1120-1128   (2008).

\bibitem{zocchi13}
D.S. Sanchez, H. Qu, D. Bulla, G. Zocchi,  \textit{Phys. Rev. E}  \textbf{87}, 022710 (2013).

\bibitem{kim14}
T.T. Le, H.D. Kim,    \emph{Nucleic Acids Res.}   \textbf{42}, 10786-10794 ( 2014).

\bibitem{ejte15}
H. Salari, B. Eslami-Mossallan, S. Naderi, M.R. Ejtehadi, \emph{J. Chem. Phys.} \textbf{143}, 104904 (2015).

\bibitem{ejte12}
A. Fathizadeh, B. Eslami-Mossallam, M. R. Ejtehadi, \emph{Phys. Rev. E}     \textbf{86},  051907 (2012).

\bibitem{tan15} 
Y.Y. Wu, L. Bao, X. Zhang, Z.J. Tan,  \emph{J. Chem. Phys.}  \textbf{142}, 125103 ( 2015).

\bibitem{io16a}
M. Zoli, \textit{Phys. Chem. Chem. Phys.}  {\bf 18}, 17666  (2016).

\bibitem{yan15} 
P. Cong, L. Dai,  H. Chen,  J.R.C. van der Maarel,  P.S. Doyle, J. Yan, \textit{Biophys. J.}  \textbf{109}, 2338-2351 (2015).


\bibitem{dan52}
H. E. Daniels, \textit{Proc. Roy. Soc. (Edinburgh) A} \textbf{63}, 290 (1952).

\bibitem{stock1}
W. Gobush, H. Yamakawa, W.H. Stockmayer,  W.S. Magee, \textit{J. Chem. Phys.} \textbf{57}, 2839-2843 (1972).

\bibitem{frey96}
J. Wilhelm, E. Frey, \emph{Phys. Rev. Lett.} \textbf{77}, 2581-2584 (1996).


\bibitem{wink03}
R.G. Winkler, \emph{J. Chem. Phys.} \textbf{118}, 2919-2928 (2003).

\bibitem{ever10}
N.B. Becker, A. Rosa, R. Everaers, \textit{Eur. Phys. J. E} \textbf{32}, 53-69 (2010).

\bibitem{tan17}
X. Zhang, L. Bao, Y.Y. Wu, X.L. Zhu,  Z.J. Tan,   \emph{J. Chem. Phys.}  \textbf{147}, 054901 (2017).

\bibitem{menon13}
R. Padinhateeri, G.I. Menon,   \emph{Biophys. J.}  \textbf{104},  463-471 (2013).


\bibitem{york05}
K. Range, E. Mayaan, L.J. Maher, D.M. York, \emph{Nucleic Acids Res.}  \textbf{33}, 1257-1268 (2005).

\bibitem{onuf16}
A.V. Drozdetski,  I.S. Tolokh,  L. Pollack,  N. Baker,  A.V. Onufriev, \emph{Phys. Rev. Lett.} \textbf{117}, 028101 (2016).

\bibitem{io12}
M. Zoli, \textit{J. Phys.: Condens. Matter} {\bf 24},  195103  (2012).

\bibitem{druk01}
K. Drukker, G. Wu, G.C. Schatz,   \emph{J. Chem. Phys.} \textbf{114},  579-590 (2001).


\bibitem{zhang97}
Y.L. Zhang, W.M. Zheng, J.X. Liu,  Y.Z. Chen,   \textit{ Phys. Rev. E}  \textbf{56}, 7100-7115 (1997).

\bibitem{n1}
The divergence in the partition function arises from the translational (zero) mode, all $r_i$'s equal, for which the two particle potential vanishes. Generally, in semiclassical problems, the zero eigenvalue can be extracted from the fluctuation factor [M. Zoli, \textit{J. Math. Phys.} {\bf 48}, 082101  (2007)]. Here however the one particle on-site potential breaks the translational invariance of the system.

\bibitem{ares05}
S. Ares, N.K. Voulgarakis, K.{\O}. Rasmussen,  A.R. Bishop,  \emph{Phys. Rev. Lett.} \textbf{94},  035504  (2005).

\bibitem{pey93b}
T. Dauxois, M. Peyrard, A.R. Bishop,  \emph{Phys. Rev. E}  \textbf{47},  R44-47 (1993).

\bibitem{joy08} 
S. Buyukdagli, M. Joyeux,  \textit{Phys.\ Rev.\ E} {\bf 77}, 031903 (2008).


\bibitem{io18}
M. Zoli,  \textit{Physica A} \textbf{492}, 903-915 (2018).

\bibitem{bres07} 
P.L. Privalov, A.I. Dragan, C. Crane-Robinson, K.J. Breslauer, D.P. Remeta, C. A.S.A. Minetti, \textit{J. Mol. Biol.}  \textbf{365}, 1–9 (2007).


\bibitem{krueg06}
A. Krueger, E. Protozanova, M.D. Frank-Kamenetskii,  \textit{Biophys. J.}  {\bf 90}, 3091-3099 (2006).

\bibitem{zdrav06}
S. Zdravkovi\'{c}, M.V. Satari\'{c},  \emph{Phys. Rev. E}  {\bf 73},  021905 (2006).

\bibitem{fenn08}
R. S. Mathew-Fenn, R. Das, and P. A. B. Harbury,  \textit{Science}  \textbf{322}, 446-449 (2008).

\bibitem{weber13}
G. Weber, \textit{Nucl. Acids Res.}  \textbf{41}, e30 (2013).

\bibitem{singh15}
A. Singh, N. Singh,     \textit{Physica A}  \textbf{419}, 328-334 (2015).


\bibitem{weber15}
I. Ferreira, T.D. Amarante,  G. Weber,    \emph{J. Chem. Phys.} \textbf{143}, 175101 (2015).

\bibitem{io14a}
M. Zoli,  \textit{Soft Matter} {\bf 10}, 4304-4311 (2014).

\bibitem{io09}
M. Zoli,   \emph{Phys.Rev. E}   \textbf{79},  041927 (2009).

\bibitem{io10}
M. Zoli,  \emph{Phys.Rev. E}    \textbf{81},  051910 ( 2010).

\bibitem{fehi}
R.P. Feynman,  A.R. Hibbs,   {\it Quantum Mechanics and Path Integrals}, (Mc Graw-Hill, New York,  1965).

\bibitem{jack}
R. Jackiw,  \textit{Rev. Mod. Phys.}  \textbf{49}, 681-706 (1977).

\bibitem{io11a}
M. Zoli,    \textit{Eur. Phys. J. E}   {\bf 34}, 68 (2011).


\bibitem{landau}
L.D. Landau, E.M. Lifshitz, \emph{Quantum Mechanics}, (Butterworth-Heinemann, Oxford, 1977).

\bibitem{pey04}
M. Peyrard,      \textit{Nonlinearity}  {\bf 17}, R1 (2004).

\bibitem{hand12}
A. Sulaiman, F.P. Zen,  H. Alatas,  L.T. Handoko,   \textit{Phys. Scripta} \textbf{86}, 015802 (2012). 

\bibitem{io14b} 
M. Zoli,   \textit{J. Theor. Biol.}   {\bf 354},  95-104 (2014). 

\bibitem{sung15}
J. Shin, O.-C. Lee, W. Sung,  \emph{J. Chem. Phys.}  \textbf{142}, 155101 (2015).


\bibitem{io03}
M. Zoli,   \textit{Phys. Rev. B}  {\bf 67}, 195102 (2003).

\bibitem{io05}
M. Zoli,    \emph{Phys. Rev. B} \textbf{71},  205111   (2005).

\bibitem{chers15} 
J. Shin,  A.G. Cherstvy,  R. Metzler, \textit{ACS Macro Lett.}  \textbf{4}, 202-206 (2015).

\bibitem{gueron}
M. Gu\'{e}ron, M. Kochoyan, J.L. Leroy,  \textit{Nature} \textbf{ 328}, 89-92 (1987).

\bibitem{benham99}
R.M. Fye,  C.J. Benham,  \textit{Phys. Rev. E}  {\bf 59}, 3408-3426 (1999).

\bibitem{bonnet03}
G. Altan-Bonnet,  A. Libchaber, O. Krichevsky,    \emph{Phys. Rev. Lett.}  \textbf{90}, 138101 (2003).

\bibitem{zocchi03}
Y. Zeng, A. Montrichok, G. Zocchi,     \emph{Phys. Rev. Lett.} \textbf{91},  148101 (2003).

\bibitem{rapti06}
Z. Rapti, A. Smerzi, K.{\O}. Rasmussen, A.R. Bishop, C.H. Choi, and A. Usheva,  \emph{Phys. Rev. E}   \textbf{73},   051902 (2006).



\bibitem{kalos09}
G. Kalosakas, S. Ares,     \textit{J. Chem. Phys.}   {\bf 130}, 235104 (2009).

\bibitem{kalos11} 
A. Apostolaki, G. Kalosakas,   \textit{Phys. Biol.}  \textbf{8}, 026006 (2011). 

\bibitem{singh11}
S. Srivastava, N. Singh,  \emph{J. Chem. Phys.} \textbf{134},   115102 (2011).

\bibitem{io13}
M. Zoli, \textit{J. Chem. Phys. } {\bf 138}, 205103 (2013).


\bibitem{metz10}
J.-H. Jeon, J. Adamcik,  G. Dietler,  R. Metzler,   \emph{Phys. Rev. Lett.} \textbf{105},  208101, (2010).

\bibitem{busta92}
S. Smith, L. Finzi, C. Bustamante,  \emph{Science}   \textbf{258},  1122-1126 (1992).

\bibitem{marko97}
J.F. Marko,  \emph{Europhys. Lett.}   {\bf 38}, 183-188 (1997).

\bibitem{busta06} 
J. Gore, Z. Bryant, M. N\"{o}llmann, M.U. Le, N.R. Cozzarelli and C. Bustamante, \textit{ Nature}  \textbf{442}, 836-839 (2006).

\bibitem{mameren09} 
J. van Mameren, P. Gross, G. Farge, P. Hooijman, M. Modesti, M. Falkenberg,
G.J.L. Wuite and E.J.G. Peterman, \emph{Proc. Natl. Acad. Sci. USA}  \textbf{106},  18231-18236 (2009).

\bibitem{wuite11} 
P. Gross, N. Laurens, L.B. Oddershede, U. Bockelmann, E.J.G. Peterman and G.J.L. Wuite,    \textit{ Nature Phys.}  \textbf{7}, 731–736 (2011).

\bibitem{marko15}
J.F. Marko,  \textit{Physica A}  \textbf{418}, 126-153 (2015).

\bibitem{io17}
M. Zoli, \textit{J. Phys.: Condens. Matter} {\bf 29},  225101  (2017).


\bibitem{wang79}
J.C. Wang,   \emph{Proc. Natl. Acad. Sci. USA}   \textbf{76},  200-203 (1979).

\bibitem{duguet93}
M. Duguet,  \emph{Nucleic Acids Res.}  \textbf{21}, 463-468 (1993).

\bibitem{bohr11}
K. Olsen, J. Bohr, AIP Advances \textbf{1}, 012108 (2011).

\bibitem{n3}
For a short chain, the molecule axis is planar as the energetic cost to bend the backbone
out of the plane is high. Accordingly, if the short chain closes into a ring, the supercoiling is mostly partitioned into
twisting while the writhing vanishes. See e.g., 
A.D. Bates, A. Maxwell,   \emph{DNA Topology} (Oxford University Press, Oxford, 2009).
 

\bibitem{n4}
This can be easily checked numerically by noticing that for, a freely jointed chain with $N$ bonds of equal length $d$, the Fourier transform of the end-to-end distribution is a product of $N$ Fourier transform of the single bond probability i.e., $P(\textbf{x})=\, (4\pi d^2)^{-1}\delta (|\textbf{x}| - d)$.

\bibitem{kp}
0. Kratky, G. Porod,  \textit{Recl. Trav. Chim. Pays Bas.}   \textbf{68}, 1106-1122 (1949).

\bibitem{croq99} 
C. Bouchiat, M.D. Wang,  J.-F. Allemand,  T. Strick,  S. M. Block,  V. Croquette, \textit{Biophys. J.}  76, 409-413 (1999).

\bibitem{odi}
T. Odijk, \textit{J. Polym. Sci., Polym. Phys. Ed.} \textbf{15}, 477 (1977).


\bibitem{skofix}
J. Skolnick, M. Fixman, \textit{Macromolecules} \textbf{10}, 944 (1977).

\bibitem{reed91}
C.E. Reed, W.F. Reed, \emph{J. Chem. Phys.}  \textbf{94}, 8479 (1991).

\bibitem{thirum95}
B.-Y. Ha, D. Thirumalai, \emph{Macromolecules} \textbf{28}, 577 (1995).

\bibitem{soder97} 
M. Ullner, B. J\"{o}nsson, C. Peterson, O. Sommelius,  B. S\"{o}derberg, \emph{J. Chem. Phys.}   \textbf{107}, 1279 (1997).

\bibitem{manni06}
G.S. Manning,  \textit{Biophys. J.}  \textbf{91}, 3607 (2006).


\bibitem{cifra10} 
P. Cifra, Z. Benkov\'{a}, T. Bleha,  \textit{Phys. Chem. Chem. Phys.}  \textbf{ 12}, 8934-8942 (2010).

\bibitem{volo11}
S. Geggier, A. Kotlyar, A. Vologodskii,   \emph{Nucl. Acids Res.}   \textbf{39}, 1419-1426 (2011).

\bibitem{save12} 
A. Savelyev, \textit{Phys. Chem. Chem. Phys.}  \textbf{ 14}, 2250-2254 (2012).

\bibitem{mazur14}
A.K. Mazur, M. Maaloum,    \emph{Phys. Rev. Lett.}   \textbf{112},  068104 (2014).

\bibitem{manghi15} 
A. Brunet, C. Tardin, L. Salom\'{e}, P. Rousseau, N. Destainville, and M. Manghi, \emph{Macromolecules}  \textbf{48}, 3641-3652 (2015).

\bibitem{chers11} 
A. G. Cherstvy, \textit{J. Phys. Chem. B}  \textbf{115}, 4286-4294 (2011).

\bibitem{archer06} 
C. Yuan,  E. Rhoades,  X.W. Lou,  L.A. Archer,  \emph{Nucl. Acids Res.}  \textbf{ 34},  4554-4560   (2006).
















































\end{thebibliography}
\end{document}